\documentclass[twoside]{article}

\usepackage{lipsum} 

\usepackage[sc]{mathpazo} 
\usepackage[T1]{fontenc} 
\usepackage{microtype} 

\usepackage[hmarginratio=1:1,top=32mm,columnsep=20pt]{geometry} 
\usepackage{multicol} 
\usepackage[hang, small,labelfont=bf,up,textfont=it,up]{caption} 
\usepackage{booktabs} 
\usepackage{float} 
\usepackage{hyperref} 

\usepackage[T1]{fontenc}
\usepackage[polish,english]{babel}
\usepackage[utf8]{inputenc}
\usepackage{lmodern}
\usepackage{verbatim}
\usepackage{amsthm}
\usepackage{graphicx} 

\usepackage{algorithm2e}

\usepackage{lettrine} 
\usepackage{paralist} 

\usepackage{abstract} 

\usepackage{titlesec} 
\renewcommand\thesection{\Roman{section}} 
\renewcommand\thesubsection{\Roman{subsection}} 
\titleformat{\section}[block]{\large\scshape\centering}{\thesection.}{1em}{} 
\titleformat{\subsection}[block]{\large}{\thesubsection.}{1em}{} 

\usepackage{fancyhdr} 
\newtheorem{tw}{Theorem}
\newtheorem{lem1}{Lemma}
\newtheorem{fac1}{Fact}

\title{\vspace{-15mm}\fontsize{24pt}{10pt}\selectfont\textbf{4/3 Rectangle Tiling lower bound}\footnote{This work was supported by the Polish National Science Centre grant
DEC-2012/06/M/ST6/00459.}} 

\author{
\large
\textsc{Grzegorz Głuch, Krzysztof Loryś}
\\[2mm] 
\normalsize University of Wrocław \\ 
\vspace{-5mm}
}
\date{}

\begin{document}

\maketitle 

\thispagestyle{fancy} 


\begin{abstract}

The problem that we consider is the following: given an $n \times n$ array $A$ of positive numbers, find a tiling
using at most $p$ rectangles (which means that each array element must be covered by some rectangle and no two
rectangles must overlap) that minimizes the maximum weight of any rectangle (the weight of a rectangle is the sum
of elements which are covered by it). We prove that it is NP-hard to approximate this problem to within a factor
of \textbf{1$\frac{1}{3}$} (the previous best result was $1\frac{1}{4}$).

\end{abstract}

\bigskip


\begin{multicols}{2} 

\section{Introduction}

\textbf{RTILE problem.}
Given an $n \times n$ array $A$ of positive numbers, find a tiling
using at most $p$ rectangles (that is rectangles which cover $A$ without overlap) that minimizes the maximum weight of any rectangle (the weight of a rectangle is the sum of elements which are covered by it). A tile is any rectangular subarray of A.\\

\textbf{Previous work.}
The problem was considered by Khanna, Muthukrishnan and Paterson in \cite{KMP98} where a 2$\frac{1}{2}$ upper bound was given. Next it was improved, firstly to 2$\frac{1}{3}$, independently by Sharp in \cite{S99} and by Loryś, Paluch in \cite{LP00}, and later to 2$\frac{1}{5}$ by Berman, DasGupta, Muthukrishnan and Ramaswami in \cite{BDMR01}. The best known result for the upper bound is 2$\frac{1}{8}$ and was presented by Paluch in \cite{P04}. The only known lower bound, equal to 1$\frac{1}{4}$, was given in \cite{KMP98}. 

In this paper we obtain a 1$\frac{1}{3}$ lower bound for the RTILE problem. As the core of the proof we use a modified construction of the one used in \cite{KMP98}. 

The main result of the paper is the following theorem.

\begin{tw}
The RTILE problem is NP-hard, even in the case where the element weights are integers in the range $[1,3]$.
Furthermore it is NP-hard to determine the optimal value to within a factor of 1$\frac{1}{3}$.
\end{tw}

\section{Proof of the Theorem}

In the proof we reduce \textsf{PLANAR-3SAT} into \textsf{RTILE}. It was shown in \cite{L82}
that \textsf{PLANAR-3SAT} is NP-complete. An instane of \textsf{PLANAR-3SAT} problem is a 3CNF formula $F$ with
an extra property that the following graph $G_{F}$ is planar. The bipartite graph $G_{F}$ has variables and clauses
as its two vertex sets. An edge $(x,c)$ exists in $G_{F}$ if and only if $x$ or $\neg x$ occurs in clause $c$.

For any formula $F$, an instance of \textsf{PLANAR-3SAT} problem, we will construct an array $A_{F}$ of integers 
in the range [1,3] and an integer $p_{F}$, which will give us an instance $(A_{F},p_{F},3)$ of \textsf{RTILE} problem.
We will construct this instance in such a manner that $F$ is satisfiable if and only if $A_{F}$ can be tiled by $p_{F}$
tiles of weight at most $3$.

\subsection{Intuitions}

Given $F$ which is an instance of \textsf{PLANAR-3SAT} we want to construct an array $A_{F}$. For every
variable in $F$ we want to create a closed loop of orthogonally adjacent fields so that every two loops are disjoint
(they don't even touch each other). We also want to place these loops on the $A_{F}$ in such a way that for every clause
$C$ in $F$, the three loops corresponding to variables occuring in $C$ meet in a special \textit{clause gadget} (Fig. 1).
\bigskip

\begingroup
    \centering
    \includegraphics[width = \linewidth]{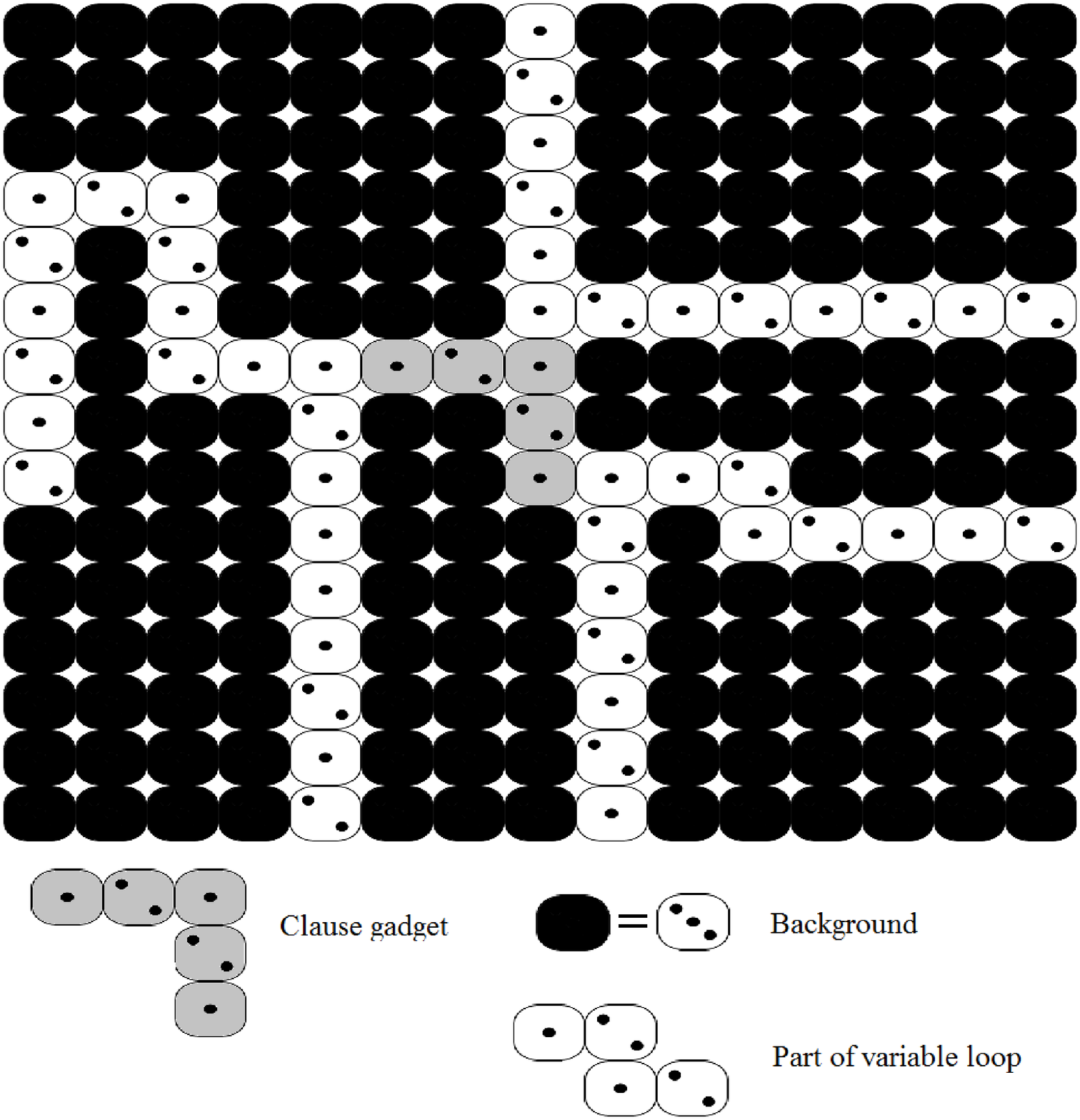}
    \captionof{figure}{Clause gadget and variable loops}\label{fig:a}
\endgroup
\bigskip
Then every loop is filled with numbers in such a way that it is possible to cover the loop in only two different ways (Fig. 2, Fig. 3) (we will call them modes) which corresponds to a valuation of a variable. \bigskip

\begingroup
    \centering
    \includegraphics[width = \linewidth]{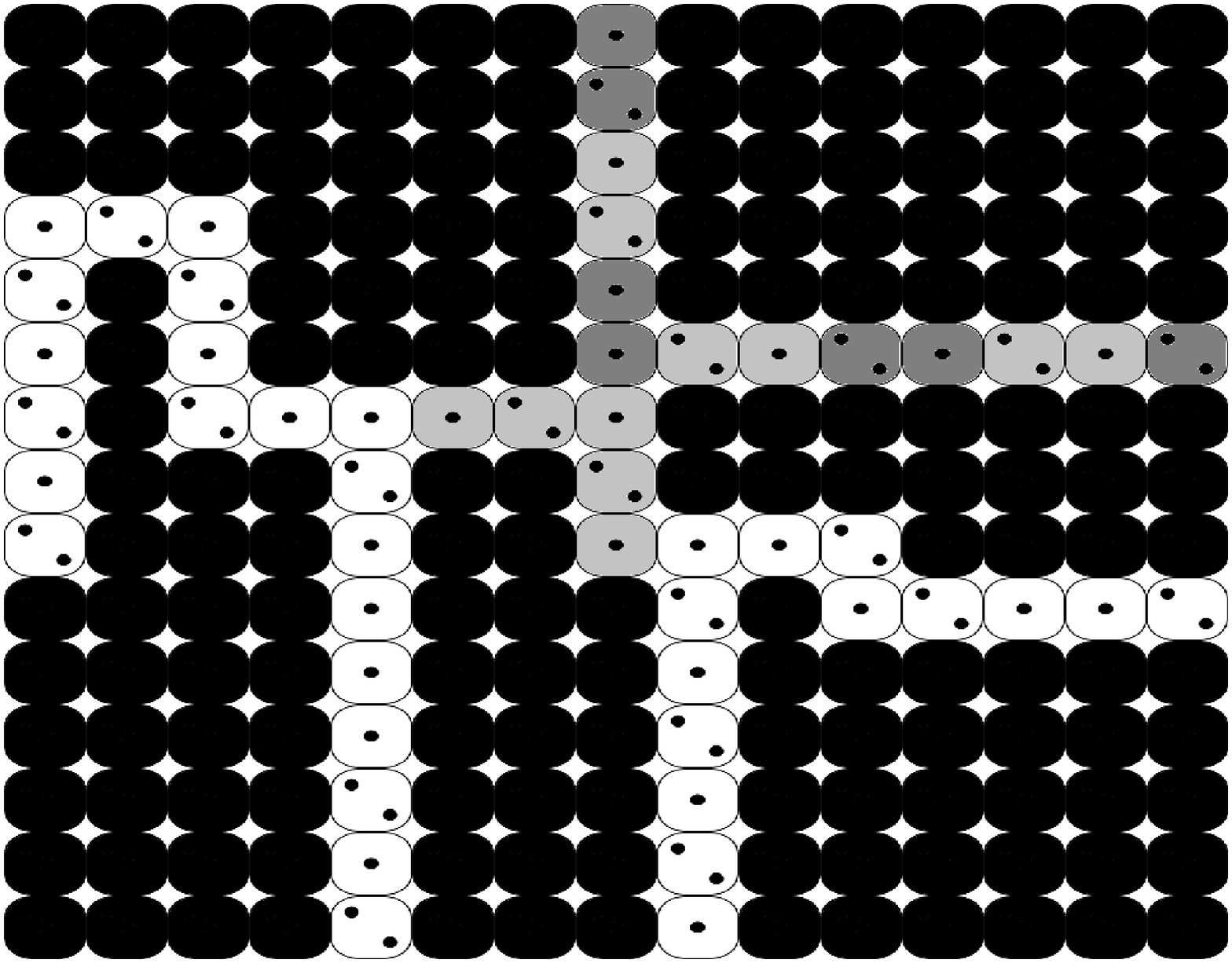}
    \captionof{figure}{Loop covering 1}\label{fig:b}
\endgroup
\bigskip

As we can see in most cases one tile covers only two consecutive elements of a variable loop. What is more if a loop is covered in one mode then part of the clause gadget (namely one of its 1's) can be joined to a tile from the loop (as in the Fig 2.) but it is impossible when a loop is in the other mode (as in the Fig 3.). Situation in the Fig 2. corresponds to a fact that under this valuation that variable causes the clause to evaluate to true. Later we will say that a given covering mode of a loop is in a good phase in proximity of a gadget clause if it is possible to join a 1 from this gadget to a tile from the loop. Finally it is necessary to use three tiles to cover clause gadget if none of its 1's is joined with a variable loop and only two tiles when one or more of its 1's is joined with a loop. 
\bigskip

\begingroup
    \centering
    \includegraphics[width = \linewidth]{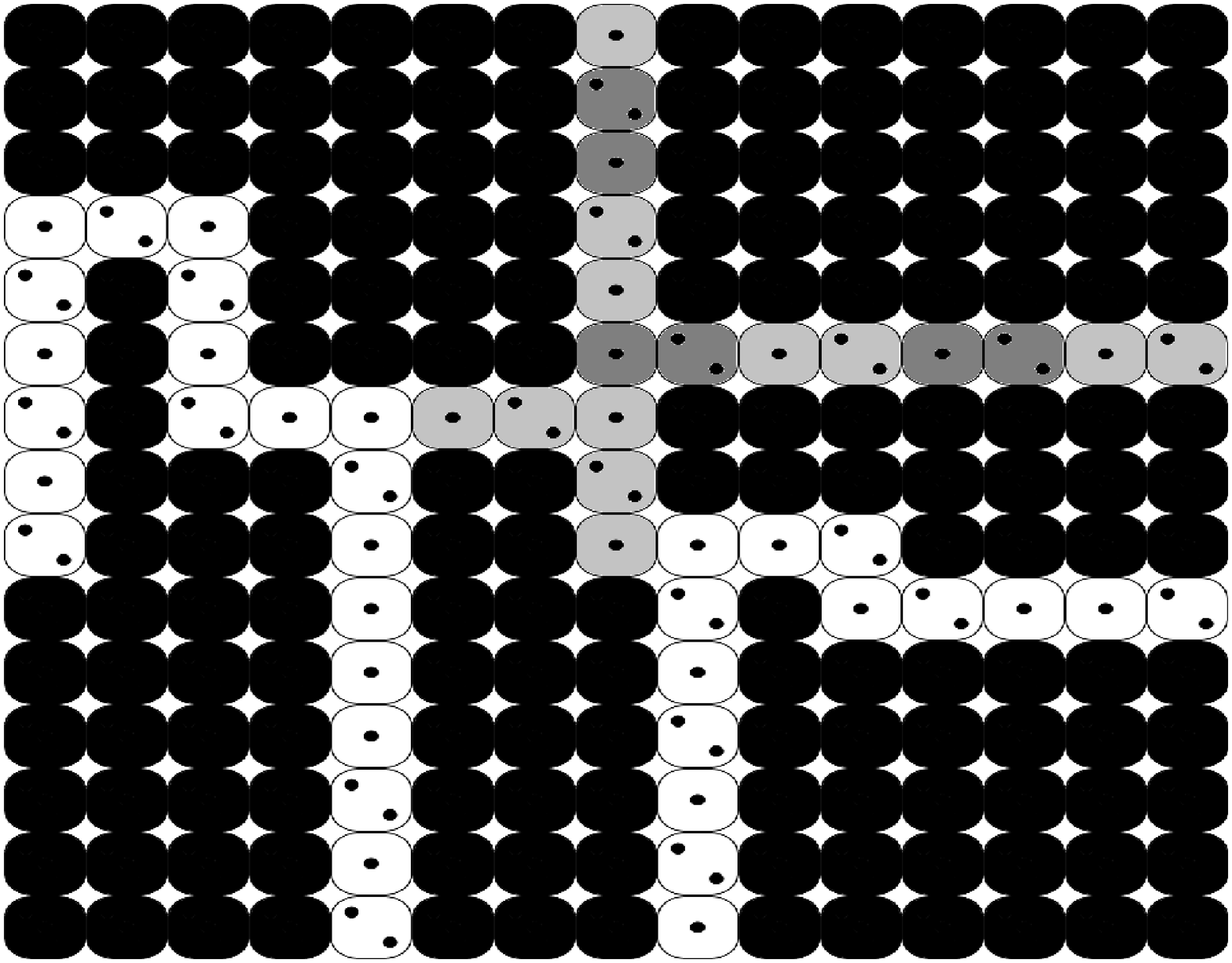}
    \captionof{figure}{Loop covering 2}\label{fig:c}
\endgroup

\subsection{Construction}

At first it is possible to construct a topology described above because of the Schnyder theorem \cite{S90} which states that every
graph with $n \geq 3$ vertices has a straight line embedding on the $n-2$ by $n-2$ grid so that each vertex lies on a grid point.

Next we need to show how to choose values of $A_{F}$ elements. The background is simply filled in with 3's while each clause gadget is filled as can be seen at Fig. 1. Next the exact filling of variable loops proceed as follows. Each variable loop is filled independently, so let's consider one of them and call it $Z$. At first we fill neighbourhoods of clause gadgets of $Z$ loop with a sequence 2112 in such a manner that two 1's from 2112 sequence and a 1 from clause gadget form a 3x1 rectangle (Fig. 1). Next we traverse the loop clockwise filling it with alternating 1's and 2's. When filling the loop we may sometimes need to use a sequence 2112 to avoid two consecutive 2's in the loop. What is more we need to provide that being in a good phase near one gadget implies being in a good (or bad) phase in proximity of another gadget depending on whether the corresponding variable occurs in a clause with a negation or not. To assure that property we use (or not) a sequence 211112 (called \textit{phase changer}) between consecutive clause gadgets.

One last thing is the choice of $p_{F}$. Denoting the number of variables as $n$, the number of clauses as $k$, the number of elements with value 3 as $t$, clause gadget as $C_{j}$, loop corresponding to a variable $x_{i}$ as $Z_{i}$, length of a variable loop $Z_{i}$ as $|Z_{i}|$ and the number of \textit{phase changers} in $Z_{i}$ as $P(Z_{i})$ we define $p_{F} = \frac{1}{2} (\sum_{i=1}^{n} ( |Z_{i}| - P(Z_{i}) ) ) + 2k + t$.

\subsection{Correctness of the Construction}

\begin{fac1}
Every closed loop on a rectangle grid has even length.
\end{fac1}

Proof. We can treat the array $A_{F}$ as a bipartite graph $G$ were vertices correspond to elements of $A_{F}$ and vertices are connected when corresponding elements are orthogonally adjacent. We can see then that a loop corresponds to a cycle in $G$ and because of that is of an even length. \qed

\begin{fac1}
For every variable loop $Z_{i}$ the number of \textit{phase changers} $P(Z_{i})$ is even.
\end{fac1}

Proof. Firstly let us notice that a decision of placing each phase changer was made solely based on the length of a variable loop between two consecutive clause gadgets and negations of the corresponding variable in two appropriate clauses. A phase changer was placed in a segment when the length of that segment was even and negations were different or when the length was odd and negations were the same. Let $Diff$ be the number of segments between two consecutive clause gadgets where the corresponding variable appeared with different negations and $Odd$ be the number of segments of odd lengths. Both values: $Odd$ and $Diff$ are even, which follows from the Fact 1. and the fact that every variable loop is a cycle. Finally the number of phase changers placed is equal to $Odd - Diff + 2x$ (where $x$ is the number of segments of even lengths with different negations at the ends). \qed

\begin{lem1}
$p_{F}$ is an integer.
\end{lem1}

Proof. It follows easily from Fact 1 and Fact 2. \qed

\begin{tw}
$F$ is satisfiable if and only if $A_{F}$ can be tiled by $p_{F}$ tiles of weight at most $3$.
\end{tw}

Proof. Implication to the right is obvious, we translate valuations of variables into tiling modes of variable loops getting a valid tiling.

Now we will prove an implication to the left.

\begin{lem1}
Every variable loop $Z_{i}$ is covered by at least $\frac{1}{2}(|Z_{i}| - P(Z_{i}))$ tiles.
\end{lem1}

Proof. We prove it by contradiction. Let $a_{i}$ be the number of tiles which covers exactly $i$ elements of a variable loop $Z_{i}$. Then we have a series of (in)equalities: 
\begin{itemize}
\item{$a_{1} + a_{2} + a_{3} < \frac{1}{2}(|Z_{i}| - P(Z_{i}))$}
\item{$a_{3} \leq P(Z_{i})$}
\item{$a_{1} + 2*a_{2} + 3*a_{3} = |Z_{i}|$} 
\end{itemize}
After solving this set of (in)equalities we get a contradiction. \qed

\begin{fac1}
Every $C_{j}$ is covered by at least $2$ tiles disjoint with those covering variable loops.
\end{fac1}

Proof. That is an obvious fact already mentioned in 'Intuitions' section. \qed

Lemma 2, Fact 3 and construction of $p_{F}$ give us that we need to use exactly $\frac{1}{2}(|Z_{i}| - P(Z_{i}))$
tiles for covering a variable loop $Z_{i}$. That means that every $Z_{i}$ is covered by $P(Z_{i})$ tiles covering three elements of a loop and $\frac{1}{2}(|Z_{i}| - 3*P(Z_{i}))$ tiles covering two loop elements. Then we see that choosing one tile of a loop determines the whole tilling of that loop, or in other words it determines a covering mode. It comes from the fact that the only places where one tile can cover more than two elements of a loop are in \textit{phase changers}.


Now we want to translate a tiling of a variable loop $Z_{i}$ using $\frac{1}{2}(|Z_{i}| - P(Z_{i}))$ tiles into a valuation of a corresponding variable $x_{i}$. If there is no clause gadget such that one of its 1's is joined to the loop $Z_{i}$ then the valuation of $x_{i}$ can be chosen arbitrary. So let's say a 1 from a clause gadget $C_{j}$ is attached to the loop $Z_{i}$. Then we choose the valuation of $x_{i}$, so that the literal in which $x_{i}$ occurs in $C_{j}$ evaluates to true. 

\begin{lem1}
Choice of valuation of $x_{i}$ is consistent among $Z_{i}$. It means that whenever some 1 from a clause gadget $C_{j'}$ is attached to the loop $Z_{i}$ then $x_{i}$ appears in $C_{j'}$ with the same "negation" as in $C_{j}$.
\end{lem1}

Proof. The 1 attached to $Z_{i}$ near $C_{j}$ forces the loop to be covered as in the Fig 2. in proximity of $C_{j}$, which in turn forces the covering mode of $Z_{i}$. Now let's consider next clause gadget after $C_{j}$ in clockwise order (namely $C_{j+1}$). We assured during the construction that being in a good phase near one clause gadget forces being in a good/bad phase in proximity of the next clause gadget. That means that if the variable $x_{i}$ occurs in the $C_{j+1}$ with the same "negation" as in $C_{j}$ (clauses look both like $...{\vee}x_{i}{\vee}...$ or like $...{\vee}{\neg}x_{i}{\vee}...$) then $Z_{i}$ is also in a good phase near $C_{j+1}$. If $x_{i}$ occurs with different "negations" in $C_{j}$ and $C_{j+1}$ then $Z_{i}$ is in a bad phase near $C_{j+1}$. By easy inductive argument we get that it is only possible for some other 1 from some other clause gadget $C_{l}$ to be attached to the $Z_{i}$ loop when $x_{i}$ occurs in $C_{l}$ with the same negation as in $C_{i}$. \qed

In this manner we constructed a valuation that satisfies F, what concludes the implication to the left. \qed

\subsection{Conclusions}

As we can see the problem of finding RTILE lower bound is far from being solved. Even the existing techniques when applied in slightly more rigorous and careful manner can lead to better results. We also hope that our result will stimulate further research in this area.



\end{multicols}


\begin{thebibliography}{99} 

\bibitem[KMP98]{KMP98}
S. Khanna, S. Muthukrishnan, M. Paterson.
On Approximating Rectangle Tiling and Packing.
{\em Soda 1998}.

\bibitem[P04]{P04}
K. Paluch.
A 17/8-Approximation Algorithm for Rectangle Tiling. 
{\em ICALP 2004 1054-1065}.

\bibitem[S99]{S99}
J. Sharp.
Tiling Multi-dimensional Arrays.
{\em Proc. 12th FCT, Springer, 1999, LNCS 1684, 500-511}.

\bibitem[LP00]{LP00}
K. Loryś, K. Paluch.
Rectangle Tiling.
{\em Proc. 3rd APPROX, Springer, 2000, LNCS 1923, 206-213}.

\bibitem[BDMR01]{BDMR01}
P. Berman, B. DasGupta, S. Muthukrishnan, S. Ramaswami.
Efficient Approximation Algorithms for Tiling and Packing Problems with Rectangles.
{\em J. Algorithms 41(2): 443-470 (2001)}.

\bibitem[L82]{L82}
D. Lichtenstein.
Planar formulae and their uses.
{\em SIAM J. Computing 11, 329-343, 1982}.

\bibitem [S90]{S90}
W. Schnyder.
Embedding Planar Graphs on the Grid.
{\em SODA 1990, 138-148}.
\end{thebibliography}
\end{document}